# Negative activation energy and dielectric signatures of excitons and excitonic Mott transitions in quantum confined laser structures

**Amit Bhunia[1], Kanika Bansal[1], Mohamed Henini[2], Marzook S. Alshammari[3], and Shouvik Datta[1, *]**

[1]Department of Physics, Indian Institute of Science Education and Research, Pune 411008, Maharashtra, India

[2]School of Physics and Astronomy, and Nanotechnology and Nanoscience Center, University of Nottingham, Nottingham NG7 2RD, UK

[3]The National Center of Nanotechnology, KACST, Riyadh 11442, Saudi Arabia

***Corresponding author's email:** shouvik@iiserpune.ac.in

# ABSTRACT

Mostly optical spectroscopies are used to investigate the physics of excitons, whereas their electrical evidences are hardly explored. Here we examined forward bias activated differential capacitance response of GaInP/AlGaInP based multi-quantum well laser diodes to trace the presence of excitons using electrical measurements. Occurrence of 'negative activation energy' after light emission is understood as thermodynamical signature of steady state excitonic population under intermediate range of carrier injections. Similar corroborative results are also observed in InGaAs/GaAs quantum dot laser structure grown by molecular beam epitaxy. With increasing biases, measured differential capacitance response slowly vanishes. This represents gradual Mott transition of excitonic phase into electron-hole plasma in GaInP/AlGaInP laser diode. This is further substantiated by more and more exponentially looking shapes of high energy tails in electroluminescence spectra with increasing forward bias, which originates from a growing non-degenerate population of free electrons and holes. Such experimental correlation between electrical and optical properties of excitons can be used to advance the next generation excitonic devices.

# I. INTRODUCTION

When a material is either optically or electrically excited, negatively charged electrons and positively charged holes can pair together under mutual Coulomb attraction to form 'positronium' like bound quasiparticle states, called excitons[1-4]. At low charge carrier densities, exciton as a composite boson of two fermions ($e^-$ and $h^+$), can either move freely or form localized complexes inside a solid. At higher densities and low enough temperatures, excitons are also expected to form condensed phases of matter like Bose Einstein Condensation (BEC) of excitons[5-9] and exciton-polaritons[10-14], BCS like superconducting states with excitonic pair formations in momentum space[15,16] and electron–hole liquid[17,18]. Many of these phenomena are finding applications in novel devices like ultralow threshold exciton-polariton lasers[19-22]. These excitons also undergo excitonic Mott transition[23] from neutral excitonic phase to conducting electron-hole plasma (EHP) phase at higher densities. However, even at room temperature, formation of exciton[24-28] can influence optical spectra of semiconductor quantum structures.

To understand these fascinating physics of excitons, a variety of optical investigations have been employed in literature. In contrast to widely reported optical studies, there are only few[29] investigations on corresponding electrical signatures of the condensed matter physics of excitonic phenomena. Being charge neutral quasiparticles, excitons cannot contribute to direct current flow, but their inherent dipolar nature makes them capable of responding to dielectric measurements. Also, excitons do not interact with defects levels in a way similar to isolated charge carriers. In this work, we exploit this characteristic difference in dielectric response of excitons as compared to free carriers using differential capacitance measurements under forward bias. We detect the existence of excitons by measuring steady state, small signal, junction capacitance response in frequency domain as carriers are injected gradually by increasing forward bias in the following III–V based electroluminescent diodes - GaInP/AlGaInP based strained multi quantum well (MQW) laser and InGaAs/GaAs based quantum dot (QD) laser.

# II. EXPERIMENTAL FRAMEWORK – PLAN, BACKGROUNDS AND INSTRUMENTATION

## A. Studying excitons using electrical measurements.

Generation of excitonic bound states and electronic transitions between discrete excitonic levels are usually accessed with optical frequencies ranging from visible to THz. However, excitons in our case are not optically generated. Here, excitons are created due to Coulomb attraction between electrons and holes which are electrically pumped within the narrow quantum well region. We further elaborate how trapping and emission of charge carriers by defect states are affected when injected carriers form excitonic bound states under high forward bias.

Instead of standard temperature activation of rate processes, here, we monitor transition rates activated by increasing forward bias at constant temperatures. Bias activated electrical response was modelled with a phenomenological rate equation which matches well with our experimental data. Normal bias activation behavior with increasing charge injection inverts after light emission and exhibits 'negative activation energy' around some intermediate injection levels. We argue that this can be attributed to the presence of a 'stable' steady state population of transitional bound states having estimated binding energy similar to those of weakly confined excitons in GaInP/AlGaInP MQW laser diodes. This thermodynamical estimation of exciton binding energy is repeatedly reproduced and further verified by qualitatively similar results for InGaAs/GaAs QD laser diodes. We also describe how one can identify subsequent Mott transition of these excitonic phases gradually into EHP at higher injection levels, using these electrical measurements. As a whole, the current investigation presents a unique and hitherto unexplored approach to study multifaceted nature of physics of excitons by looking at its electrical signatures. One can extend this understanding towards efficient electrical manipulation of excitonic processes. Therefore, this study can prove to be helpful in improving exciton based next generation devices like excitonic and polaritonic ultralow threshold lasers, excitonic solar cells etc.

**B. Basics of differential capacitance technique and activation of electronic transition rates in laser diodes under forward bias.**

Earlier, we explored[30-32] the connection between voltage modulated electroluminescence (VMEL) and electrical properties of GaInP/AlGaInP based MQW lasers. Under low frequency modulation, we demonstrated the mutual non-exclusivity of radiative and non-radiative transitions. This originates when slowly responding (~ms) sub-band gap electronic states can directly influence faster (~ps-ns) radiative recombinations in a non-trivial fashion which necessitates[30] $\frac{1}{\tau_{Effective}} \neq \frac{1}{\tau_{Non-Radiative}} + \frac{1}{\tau_{Radiative}}$ , where $\tau_{Effective}$ is the effective time constant for such events. For any applied modulation frequency ($f$) at a temperature (T), charge carriers from certain sub-band gap channels with activation energy can also contribute[30] to the total available charge carrier reservoir at the junction. This affects both junction capacitance (C) and modulated electroluminescence (VMEL) simultaneously. To fully understand the role of sub-band gap states, Fermi levels and electronic transition rates $R = \frac{1}{\tau} = \nu \exp\left(-\frac{E_{Th}}{k_B T}\right)$ , we suggest readers to see the band diagram given in Fig.3 in Ref. 30. Here $E_{Th}$ is thermal activation energy, ν is the thermal prefactor representing the heat bath, T is temperature in degree Kelvin, τ is the time period of the transition and $k_B$ is the Boltzmann constant. This band diagram was further used[32] to explain the observed anomalous temperature dependence of VMEL and negative capacitance. In this work, we extend these results to electrically probe the presence of excitons during electroluminescence (EL).

When the quasi-Fermi level crosses the activation energy ($E_{Th}$) of any charge transfer process, one observes a maximum in $G/2\pi f$ (G=conductance) and an inflection point in capacitance (C) in both frequency and temperature domains. Such resonant conductance activation in frequency domain is related to dielectric environment being probed by impedance spectroscopy[33]. To fully appreciate the above statements, we recommend the readers to consult Fig.1 in section 2 of Ref. 33 and discussions around equations (2.6), (2.8). At low frequencies, the displacement vector within the active junction can follow the applied electric field modulation. However, at high frequencies dipoles cannot respond fast enough to applied modulation. As a result, energy loss through conductance activation is maximum at some intermediate frequencies where dipolar environments respond resonantly. This constitutes a general understanding of any dielectric response measurement. We, however, emphasize that such understanding is in no way dependent on depletion approximations used only for reverse biased junctions.

However, with a derivative technique[34,35], we use plots of $f$dC/d$f$ with $f$, where it is easier to recognize such peak value of conductance activation and locate the frequency ($f_{Max}$) of peak maximum. Corresponding number density $N(E_{Th})$ which adds to number of carriers available[30] for radiative recombination is given by a generalized equation[34, 35],

$$N(E_{Th}) \approx -\left( f \frac{dC}{df} \right) \frac{U}{k_B T} \frac{1}{qw} \qquad (1)$$

where U is the effective built-in-potential and is linearly related to applied bias $V_{dc}$ such that U = [$\Phi_B$ - $V_{dc}$] and $\Phi_B$ is the built-in potential at zero bias, $q$ is the electronic charge, $w$ is the width of the junction. In case of impedance spectroscopy, the usual thermal activation energy is obtained from the slope of an Arrhenius plot of ln($f_{Max}$) with 1000/$T$. However, in this work, we are particularly interested in exploring frequency response of electrical impedance as activated by increasing voltage bias $V_{dc}$ in the active region of these light emitting diodes at constant temperatures. The estimated $fdC/df$ plays a role similar to electric modulus in such dielectric measurements. Such description using electric modulus often removes unwanted artifacts from the analysis of dielectric response measurements. In our earlier works[30], we have clearly shown how a region of diffusion capacitance under increasing forward bias evolves into a non-standard region of 'negative capacitance' with the onset of voltage modulated electroluminescence. It is important to note[30] that frequency dependence of this negative capacitance is not similar to that of an ordinary inductance, which rules out the presence of any external parasitic inductance. Moreover, we will neither use the above equation (1) in its entirety nor the standard 1/$C^2$ vs V plots to estimate carrier densities, both of which require strict implementation of depletion approximation. Currently, the above proportionality of dipolar density with estimated *fdC/df* values is used only as a qualitative measure. Any dissipative process also causes decay. Hence, the width of such *fdC/df* based conductance activation peaks in frequency domain is always related[33] to exponential decay rates of such dissipative processes in time domain. VMEL results are, however, not presented in this manuscript to avoid complexity.

### C. Instrumentation and samples.

For the steady state, small signal capacitance measurement, we used Agilent's precision LCR Meter E4980A at $f \leq$ 2MHz using 30mV rms voltage modulation. The instrument uses the null technique of an ac bridge. Measured capacitance is dependent on small signal current which is in quadrature with the applied voltage modulation. This capacitance is estimated[30] using a simple, equivalent parallel RC circuit for the laser diode under forward bias. Bias voltage (current) being applied to the device is represented as $V_{dc}$ ($I_{dc}$) and monitored by the LCR meter. For temperature variation and control, we used a CS-204S-DMX-20 closed cycle cryostat from Advance Research Systems along with the Lakeshore Model-340 temperature controller at an accuracy of ±0.1 K. We measured standard EL spectra using FDS010 Si detector and SR850 Lock-in amplifier.

The samples which have been used are - (i) GaInP/AlGaInP based strained multi quantum well (MQW) edge emitting laser diodes from Sanyo (DL 3148-025), and (ii) InGaAs/GaAs based quantum dot (QD) laser diode grown by Molecular Beam Epitaxy (MBE) on (311B) GaAs substrates. A 0.7μm-thick GaAs buffer layer was grown before deposition of $In_{0.5}Ga_{0.5}As$ strained quantum dots at 450 °C which are embedded in the center of a GaAs QW surrounded by $Al_{0.6}Ga_{0.4}As$ cladding layers on both sides (further details can be found in Ref. 36). These dots have an average diameter around 26 nm and height 1 nm. Light is emitted through a small mesa etched on its top surface.

To derive the *fdC/df*, presented in this paper, measured variation of capacitance with frequency was numerically differentiated and multiplied by the frequency itself. Savitzky-Golay signal processing algorithm using Origin-8.5 was employed to smoothen out numerically estimated *fdC/df vs f* plots.

## III. EXPERIMENTAL RESULTS AND ANALYSES

### A. Use of voltage activated rate equation.

Figures 1a and 1b demonstrate the variation of *f*dC/d*f* with frequency at two different ranges of forward bias at room temperature for GaInP/AlGaInP MQW laser diode. We obtain these peak frequencies ($f_{Max}$) from the measured *f*dC/d*f* data at different $V_{dc}$ as shown in Fig.1a. Estimated $f_{Max}$ values initially decrease when forward bias ($V_{dc}$) slowly increases from zero. We must emphasize that observations of similar dynamic behavior of small signal impedance response are also reproduced[37] in non-light emitting Silicon diodes. We have mentioned that defect response can be seen in capacitance measurements when modulation frequency (*f*) range matches with the transition rate (R). Here, one usually expects that a decrease in temperature decreases the peak position of frequency response ($f_{Max}$) and increases the overall amplitude of *f*dC/d*f* peaks. Therefore, the observed phenomenological activation rate shown in Fig. 1a, with increasing $V_{dc}$, is analogous to usual variations of *f*dC/d*f* vs 1000/*T* encountered in ordinary impedance spectroscopy. In this regime of applied bias, electronic defect mediated transitions determine dynamic response of the active junction. Empirically, this indicates that the role of $V_{dc}$ is qualitatively equivalent to 1000/T. In general, electrical injection responsible for quasi Fermi level splitting is a thermally activated barrier crossing process. Moreover, quasi Fermi levels in the active junction, which are nothing but chemical potentials, will be influenced heavily by large number of injected carriers as a result of forward bias. In case of laser diodes, with increasing forward bias these levels move towards and eventually cross the band edges to reach the transparency condition necessary for lasing. Therefore, a simplistic microscopic explanation of the above mentioned equivalence can be understood in the following way. Within our experimental parameter ranges, as we increase the forward bias ($V_{dc}$), quasi Fermi levels move towards band edges while with increasing temperature (T) Fermi levels rather move towards the mid-gap in these laser diodes. At this stage, we are certainly ignoring any material specific Fermi level pinning by defects, in order to build a simplified framework of such voltage activated processes.

With this understanding, we conceive a revised phenomenological rate equation for bias activated electronic transitions rate $R^*$ at a constant temperature as:

$$R^* \approx \frac{1}{\tau} = \nu^* \exp\left( -\frac{E_a}{k_B} \eta V_{dc} \right) \tag{2}$$

where $\eta$ is a proportionality factor in units of $V^{-1}K^{-1}$ to ensure correct dimensionality and $E_a$ is activation energy for transitions activated by bias voltage. Rewriting this equation gives,

$$\ln R^* = \left( \frac{-E_a \eta}{k_B} \right) V_{dc} + \ln \nu^* \tag{3}$$

The estimated $f_{Max}$ coincides with position of maximum transition rate $R^*$ in frequency domain. Hence, plotting ln($f_{Max}$) with $V_{dc}$ should give a straight line with slope ($m$) as

$$m = \left( \frac{-E_a \eta}{k_B} \right). \tag{4}$$

In case of standard, thermally activated Arrhenius like rate process, ν* represents the attempt-to-escape frequency related to thermal bath. Its interpretation broadly remains same in the current analysis. Saturation of defect levels at the active junction is also at its minimum at $V_{dc} = 0$ and this $V_{dc} \rightarrow 0$ limit points to largest $f_{Max}$ possible for such voltage activated transitions under forward biasing conditions. ln($f_{Max}$) vs $V_{dc}$ plot shown in the inset of Fig. 1a indeed fits nicely to straight lines in support of the above phenomenological rate equation (2). Initial slope ($m_1$) is -0.04 for very low forward biases and it abruptly increases to $m_2$ = -0.19 after ~1.1 V. This slope change around 1.1 V is related with considerable changes in carrier dynamics due to the onset of significant injection currents around this current as shown in Fig. 1c. Intercept of such plots are usually related to the entropic changes (ΔS) associated with the injection process following the standard transition rate $R = \frac{1}{\tau} = \nu \exp\left( -\frac{\Delta F}{k_B T} \right) = \left[ \nu \exp\left( \frac{\Delta S}{k_B} \right) \right] \exp\left( -\frac{\Delta H}{k_B T} \right)$ where change in free energy is $\Delta F = \Delta H - T\Delta S$; $\Delta H \approx E_a$, ΔH being change in enthalpy. Therefore, the above parameter η here can also be related to free energy change under voltage activations. Observed change in slope around 1.1 V is connected with corresponding increases in the intercept and thereby an enhanced $\Delta S$ during any electronic transition. This enhancement in $\Delta S$ for $V_{dc} > 1.1$ V is likely connected with the increased number of possible microstates of charge carriers within the active junction after significant injection starts around 1.1 V as shown in Fig. 1c.

**B. Occurrence of 'negative activation energy' and its connection to the presence of excitonic bound states in GaInP/AlGaInP MQW laser diode.**

Most importantly, in Fig. 1b, we show differential capacitive response, as $f$dC/d$f$ vs $f$ plots, in an intermediate bias regime, below lasing threshold (~$V_{dc}$ = 1.5 V to < 2 V). We also begin to observe measurable EL signal within this bias range as indicated in Fig. 1c. Interestingly, in this regime, the peak frequencies of $f$dC/d$f$ shift toward higher $f$ with increasing $V_{dc}$. The resultant Arrhenius like plot of ln($f_{Max}$) with $V_{dc}$ now shows a seemingly counterintuitive positive slope ( m ~1.99) for a reasonably linear fit. Substituting this into the equation (4) then indicates the presence of 'negative activation energy' for bias activated rate processes at $V_{dc}$ ~1.5 V and above. Therefore, around these intermediate injection levels, we observe an exact opposite dynamical behavior of bias activation than what was seen in case of lower injection levels [Fig. 1a]. We must also mention here that thermally activated Boltzmann like behavior of rate processes may no longer apply at even higher levels of carrier injections. Our data also tends to show [inset of Fig. 1b] slight deviations from linearity at higher biases which is still below the typical lasing threshold (~2.2 V, ~20 mA) of these devices. In Fig. 1d, we show how the slopes of ln($f_{Max}$) vs $V_{dc}$ change with sample temperature. We will use these slopes to estimate the binding energy in the following section.

In general, activation energy is known as a free energy barrier working against any one-step transition process. It determines the fraction of carriers which can cross this free energy barrier. However, there are well known situations in physical chemistry where certain reaction rates decrease with increasing temperatures which cannot be explained by a single step crossing of the activation energy barrier. This quintessential effective 'negative activation energy' can, therefore, be explained[38-41] only by a two-step configurational process through a stable 'transitional bound state' such that low energy initial states can make transitions to final states at a much faster rate than high energy initial states. The net effective activation energy for the whole transition is given by Tolman's interpretation,

$$E_a = \langle E_{TS} \rangle - \langle E_R \rangle \quad (5)$$

where $\langle E_{TS} \rangle$ is the average thermodynamical energy of a 'stable' population of transitional bound states measured using steady state electrical responses. $\langle E_R \rangle$ is the average energy of the initial states which, for this intermediate bias range, is typically proportional to the energy of the thermal bath ($\sim k_B T$) of injected carriers. Clearly, if the first term in equation (5) is smaller than the second term, then one can get a resultant activation energy which is negative. It is certainly tempting to associate this transitional bound state with the formation of excitons having average binding energy $\langle E_{TS} \rangle$. Subsequently, it undergoes radiative recombination and disappears as shown in Fig. 2a. This situation is characteristically different from the thermal dissociation of excitons into electron and holes which are usually probed using thermal quenching of luminescence.

For better understanding, we now discuss microscopic context of the configurational free energy diagram of this composite transition processes shown in Fig. 2a. Initially, trapped charge carriers have an average thermal energy $\langle E_R \rangle \sim k_B T$ available to them at one particular temperature (T). In general, these carriers have to cross the initial activation energy barrier $E_1$ and then directly take part in light emission. This $E_1$ can be a Coulomb barrier which takes care of mutual repulsion of similarly charged carriers before these can populate the narrow active region of the quantum well. However, with increasing injections, quasi-Fermi levels move closer to the band edges. Some of these electrons-holes can now first populate this thermodynamical transition state (indicted as excitonic state) as they experience the strong Coulomb attractions within the narrow confinement of a quantum well. This configurational bound state also has a free energy barrier $\langle E_{TS} \rangle$ beyond which it finally takes part in the radiative recombinations. We understand that the thermodynamic microstates contributing to impedance spectroscopy are characteristically different before and after the formation of excitons. This situation changes again after the recombination of excitons. Corresponding free energy changes are reflected in the configurational free energy diagram in Fig. 2a. This type of free energy diagram can ensure that the necessary condition[39] $E_{-1} > E_1 + E_2$ is satisfied to have an effective 'negative activation energy' for the overall transition process (see discussions around equation (11) in Ref. 39 for a lucid schematic description of 'negative activation energy' process and its connection to our configurational free energy diagram in line with other pioneering papers[38,40,41]).

To further verify the existence of such excitonic presence around these intermediate injection levels, we combine equations (4) and (5) to get

$$E_a = -m\frac{k_B}{\eta} = \langle E_{TS} \rangle - \langle E_R \rangle \qquad (6)$$

With decreasing temperature, $\langle E_R \rangle \sim k_B T$ should decrease and hence the slope '*m*' given in equation (4) should also depend on temperature. We indeed observe a decrease in the measured slope '*m(T)*' with decreasing temperature from 296 K down to 150 K as shown in Figs. 1b and 1d and finally depicted it as a plot of *m(T)* vs temperature in Fig. 2b. Such decrease of *m(T*) was also reproduced down to 7.9 K in similar samples. Not only we see a decrease in slope within this temperature range, but also we find that frequency of peak response ($f_{Max}$) shifts to lower frequency side with decreasing temperature. To get a quantitative estimate of $\langle E_{TS} \rangle$, we replace $\langle E_R \rangle$ by energy ~$k_B T$ of thermal bath and by re-arranging terms we get

$$m(T) = -\frac{\langle E_{TS} \rangle \eta}{k_B} + \eta T = D + nT \qquad (7)$$

This suggests that a plot of this slope *m(T)* vs *T* can resemble a straight line within this range of temperature where '*D*' is the new intercept and '*n*' is the new slope. By fitting a straight line to the data shown in Fig. 2b, we obtain the value of $\eta$ from the new slope '*n*' and then estimate $\langle E_{TS} \rangle$ from the intercept '*D*'. Here we also assume that both $\eta$ and $\langle E_{TS} \rangle$ do not vary much[42] within this range of temperature such that $\frac{\Delta\langle E_{TS} \rangle}{\langle E_{TS} \rangle} << 1$ even when band gap of the material may actually changes considerably. Finally, we estimate $\langle E_{TS} \rangle$ to be 7.1±0.9 meV. This value closely matches with the binding energy of the weakly confined excitons in these MQW laser diodes[43,44]. Therefore, we connect this occurrence of an effective 'negative activation energy' to the presence of a stable, steady state population of 'excitonic bound states'. We further highlight that contribution of defects and excitons in *fdC/df* measurements can simply be differentiated by looking at Figs. 1a and 1b for GaInP/AlGaInP MQW laser diodes. An ordinary defect related transition gives rise to usual negative slopes in Arrhenius like plots as evident from Fig. 1a. However, situation changes completely [Fig. 1b] around 1.5V in case of GaInP/AlGaInP laser diode. We emphasize that this can only be interpreted in terms of influences of steady state presence of 'intermediate bound states' identified as excitons from these capacitive measurements.

Additionally, we also want to mention that excitonic binding energies are typically less than 10 meV in these quantum confined III-V semiconductor structures with high potential energy barriers. Still it is possible to see resonant excitonic[24-28] effects at room temperature ($k_BT$ ~26meV). Unlike bulk material, these barriers oppose spatial separation of electrons and holes along the confinement direction. Additionally, inherent random fluctuations of potential barrier due to uneven well thickness and alloy disorder present across the well interface may also localize these injected electrons-holes in each other's vicinity. As a result, electrons-holes can experience each other's Coulomb attractions and subsequently form excitons at room temperature[24]. Thus, under high rate of excitation (larger than the rate of radiative recombinations), these structures can exhibit a steady state population of resonant excitons at room temperature, even though the excitonic life time is very short (~ps). Strong quantum confinement also squeezes the excitonic wave function along the confinement direction to alter the excitonic binding energy to some extent. We recommend the readers to see Fig. 6 in Ref. 24, pages 11-12, Fig. 4 in Ref. 27 and discussion above equation (4) in Ref. 28 for detailed information on this issue.

In addition, we observe that bias activated *f*dC/d*f* signatures of excitonic population gradually diminish beyond ~1.5V and virtually disappear above ~1.7 V as shown in Fig. 1b. This is likely connected with excitonic Mott transition and will be explained later in this paper. We calculate the Bohr exciton radius ($a_B$) ~8.3 nm from the above estimate of excitonic binding energy for this GaInP/AlGaInP MQW. We assume that a non-degenerate population of electron-hole gas, formed after Mott transition, obeys classical Boltzmann statistics. Then we use the Debye-Huckel model of electrostatic screening to find the Mott transition density[2] as

$$n_{Mott} = (1.19)^2 \frac{\varepsilon\varepsilon_0 k_B T}{e^2 {a_B}^2} \tag{8}$$

where ε (= 11.8) and $\varepsilon_0$ are dielectric constants of GaInP and free space, respectively, $k_B$ is the Boltzmann constant and T is the temperature in Kelvin. Our calculations yield a value of $n_{Mott} = 3.5 \times 10^{17} / cm^3$ at room temperature.

To sum up, trapping and emission of charge carriers by defect states as well as their saturation under high forward bias can change when injected carriers form excitonic bound states. As a result, we see how such formation and recombination of excitons can affect the measurement of junction capacitance through changes in dielectric environment of the active junction under large forward bias.

### C. Enhanced excitonic contribution at low temperature in GaInP/AlGaInP MQW laser diode.

In Figs. 3a and 3b, we compared *fdC/df* vs *f* plots at both 294 K and 7.7 K, respectively, under partly similar range of applied injection currents. However, unlike Fig. 1a, we do not observe usual negative slopes of $\ln(f_{Max})$ *vs* $V_{dc}$ plots for any bias levels at 7.7 K. Earlier we had also seen[32] that defect mediated VMEL signal dies down at such low temperatures. All these indicate that the origin of positive slopes in $\ln(f_{Max})$ *vs* $V_{dc}$ plots with increasing injections are not related to usual trapping and emission processes from electronic defects which are probably frozen out at such low temperatures. Therefore, this only implies more dominant presence of transitional bound states like excitons at such low temperature. Moreover, we qualitatively assume that peak magnitude of *fdC/df* , in the 'negative activation energy' regime, scales with the number of excitons present within the active junction. We further notice that peak *fdC/df* values at 7.7 K do not decrease with increasing bias currents unlike those at 294 K. All these possibly signify a lack of ionization of excitons at such low temperature as will also be explained in the next section using results of Fig. 4. We also notice that the *fdC/df* peak magnitudes are somewhat smaller (~12 pF) at lower temperature of 7.7 K as compared to ~16 pF at 294 K for a similar injection current (e.g. ~1.15 μA). At one particular injection current, we infuse approximately similar number of electrons and holes in the active region. However, the radiative recombination of the excitons is significantly stronger at lower temperature. This stronger EL readily depletes the number of available dipolar excitons which are contributing to steady state dielectric measurements. Therefore, it explains the observed reduction in *fdC/df* peak magnitude at 7.7 K. Under similar current injections, these bias activated conductance peaks represented by *fdC/df* also shift toward lower frequencies at lower temperatures as expected from our rate equation model.

### D. Concurrent changes in *fdC/df* signature of excitons and EL spectra leading to electrical identification of excitonic Mott transition in GaInP/AlGaInP MQW laser diodes.

To further establish our previous assertion on electrical signature of excitonic Mott transition near room temperature as well as larger excitonic recombinations at lower temperatures, we present some optical studies in subsequent figures. This is done to compare and corroborate optical results with their electrical counterparts. We measured standard EL spectra under similar levels of carrier injections at 294 K and at 7.9 K and finally the estimated photon flux per unit area is plotted against bias currents. The calculated photon flux strongly depends upon measurement temperature for a particular injection level and clearly demonstrates typical power law behavior with increasing forward bias current. Ideally this exponent value of 1 indicates the predominant excitonic recombination[45-47] and a value of 2 or more corresponds to free carrier recombinations. Here we estimate the power law exponents as 1.11±0.01 at 7.9 K which increases to 1.59±0.08 at 294 K. Therefore, our data presented in Fig. 4 supports stronger presence of excitonic radiative recombinations in EL at 7.9 K as compared to 294 K. We further note that any intermediate value between 1 and 2 of the power law exponent indicates the coexistence of the recombinations from excitons and free electrons and holes. The above conclusions certainly support our previous analyses based on Figs. 3a and 3b. There we also observed that excitonic contribution to *fdC/df* peak magnitude does not decrease with increasing forward bias at 7.7 K unlike that of 294 K.

High energy tails of asymmetric EL spectra at 294 K look like straight lines as shown in Fig. 5a. This increasingly exponentially shaped[48,49] nature of EL spectra is also accompanied by a red shift of the EL peak as the bias increases. On the other hand, at 7.9 K, we observe exciton like sharp resonant, symmetric spectra [Fig. 5b]. Although EL emission at 294 K starts around 1.5 V [Fig. 1c], but peak energies and line widths (FWHM) of EL given in Fig. 5c are estimated only from experimentally measurable well shaped spectra of Fig. 5a for bias values ≥1.75 V. However, more and more exponentially looking high energy spectral tails indicate that ensuing Mott transition process at room temperature is also gradual[50] due to slowly increasing fraction of free electron and holes. In fact, these high energy EL tails become prominently exponential only after 0.8 mA, 1.86 V. Most interestingly, measured electrical signatures of excitonic bound state tend to slowly disappear by ~1.7 V as shown in Fig. 1b. Therefore, this continuing suppression of electrical signature of excitonic response starting after ~1.7 V, is likely connected to optical evidence of excitonic Mott transition at 294 K.

In Figs. 5c and 5d, we have compared EL peak energies and line widths at 294 K to those at 7.9 K. Larger red shifts in peak energies at 294 K with injection of charge carriers point towards ongoing band gap renormalization effects. Additionally, due to enhanced scatterings, spectral line widths also increase significantly at higher temperatures. However, we do not see any considerable changes in peak energy and line width at 7.9 K. Therefore, it implies an apparent lack of evidence of excitonic ionizations even under similar range of high injection levels at such a low temperature, which is also evident in Figs. 3b and 4. In fact, observed minor blue shifts at 7.9 K are significantly smaller than even the excitonic binding energy. Besides, we also do not observe any saturation of EL spectra in such bias ranges at these temperatures to claim any notable phase space filling effects.

Estimated line widths (FWHM~Δ) from the EL spectra at 294 K are used to calculate the approximate number of electron-hole pair density ($n_0$) in this quasi-2D sample. The following expressions[1] under zero Kelvin limit for free electron system are used to calculate the same.

$$n_0 = n_{e,h} = \frac{\nu_{C,V}\left(2m^*_{e,h}F_{e,h}\right)^{1.5}}{3\pi^2\hbar^3} \quad (9)$$

$$F_{e,h} = (2\Delta)\left[1+\left(\frac{\nu_{C,V}}{\nu_{V,C}}\right)\frac{m^*_{e,h}}{m^*_{h,e}}\right]^{-1} \quad (10)$$

where $n_{e,h}$ is the electron or hole density, $\nu_{C,V}$ is the number of equivalent minima or maxima of the conduction or valence band, respectively, with $\nu_{C,V} \sim 1$ for the Γ point in strained GaInP band structure, $m^*_{e,h}$ is the effective mass[51] of electron and hole such that $m^*_e = 0.088m_0$ and $m^*_{hh} = 0.7m_0$ respectively, $m_0$ is the free electron mass and $F_{e,h}$ is the quasi-Fermi energy level position of electron or hole measured with respect to mid-gap such that $2\Delta = F_e + F_h$ is full width at the base of each EL spectrum. When we increase bias current from 120 μA to 5 mA, corresponding values of Δ vary from ~50.2 meV to ~68.1 meV as seen in Fig. 5c. As a result, calculated electron-hole pair density changes from $\sim 3.2\times10^{18}/cm^3$ to $\sim 5\times10^{18}/cm^3$. Therefore, even these zero Kelvin estimates of number density calculated from EL spectroscopy are certainly more than the anticipated Mott transition density of $n_{Mott}=3.5\times10^{17}/cm^3$ derived from electrical *fdC/df* measurements. This evidently indicates gradual excitonic Mott transition in these quasi-2D GaInP/AlGaInP MQWs at 294 K over such bias ranges in accordance with the analysis of Fig 4.

We must reiterate that above mentioned vanishing of *fdC/df* signature [Figs. 1b and 3a] also happens in the similar range of forward biases where we also see standard optical spectroscopic signature [Fig. 5a] of excitonic Mott transition. This concurrence of both optical and electrical effects helped us to identify these corresponding changes in *fdC/df* as those originating from the effect of formation of a steady state population of excitons and their subsequent disappearance through Mott transition. Capacitance or *fdC/df* is a measure of dielectric response of the active junction and it is not surprising that this response can change when neutral excitons dissociate into conducting electron-hole plasma through a Mott transition. Moreover, electronic defects are expected to freeze at such low temperatures of ~7-8 K. These eventually stop taking part in carrier trapping and emission processes. Defect levels also saturate under such high forward biasing. We, however, still see 'negative activation energy' like signatures of *fdC/df* even at such low temperatures [Fig. 3b].

**E. Enhanced variation of *fdC/df* signature of excitons with temperature in MBE grown InGaAs/GaAs QD laser diode.**

Comparatively stronger overlap of electron-hole wave functions from all sides within the zero dimensional InGaAs/GaAs QDs leads to dominant excitonic presence in these device structures. We find convincing support of all our previous analyses on electrical signature of excitons in these QD laser diodes as mentioned below. In Fig. 6a, we observe similar plots of large *fdC/df* peaks with positive slopes for ln$(f_{Max})$ *vs* $V_{dc}$ plots in MBE grown InGaAs/GaAs QD laser samples around 294 K. Positive slopes of ln$(f_{Max})$ vs $V_{dc}$ like Arrhenius plots or 'negative activation energy' regime start at very small forward bias. However, comparing Figs. 6a and 6b, we notice that the relative peak heights of *fdC/df* reduce much more substantially with lowering of temperature as compared to similar plots of weakly confined quasi-2D GaInP/AlGaInP MQW laser sample given in Figs. 3a and 3b. As a result, this much larger variation of *fdC/df* peak heights can be explained in terms of expectedly higher radiative recombination in QD laser at 7.7 K. Such strong radiative recombinations can significantly reduce available number of excitonic populations that can contribute to steady state dielectric measurements. These *fdC/df* peaks also shift towards larger frequencies at both the temperatures. As a result, we only see positive slope in ln$(f_{Max})$ *vs* $V_{dc}$ plots for this QD laser diode even at room temperature under any applied forward bias. This is understandable as QDs are supposed to be strongly excitonic in nature and even a very small bias can create stable excitonic population displaying 'negative activation energy' like behavior. Importantly, the excitonic binding energy estimated for this $In_{0.5}Ga_{0.5}As$ QD laser diode, from similar plots of ln$(f_{Max})$ *vs* $V_{dc}$ for different temperatures with positive slopes, is found to be 6.4±0.7 meV. This agrees well with the reported values[52,53]. In addition, height of the *fdC/df* peaks continuously increases with increasing bias currents (unlike MQW sample) at all temperatures. This indicates that dipolar excitonic population sampled by our differential capacitive measurements is also increasing with increasing injections. It signifies the absence of any Mott transition in these QD lasers as opposed to that shown in Figs. 1b and 3a for the MQW lasers. It is anticipated because the influence of Coulomb screening is substantially reduced in lower dimensional systems. This happens as the probability of finding any free charge carrier, which can screen the excitonic interaction decreases exponentially within the barrier layers of any QD structure. Besides we also note that the frequency span of bias activated *fdC/df* peaks in InGaAs/GaAs QD laser is comparatively temperature independent as compared to that of GaInP/AlGaInP MQW laser.

Lastly, we want to emphasize that all these results once again affirm our claim on electrical evidence of steady state excitonic populations at higher injections in quantum confined laser structures using dielectric impedance measurements. As mentioned above, formation of more excitons with increasing bias, not only shifts the *fdC/df* peak towards higher frequencies, but also enhances its magnitude in this QD sample. This is not expected from any ordinary defect mediated carrier trapping and emission processes as observed in Fig. 1a. The *fdC/df* amplitudes also decrease significantly at low temperatures where enhanced radiative recombinations irreversibly remove large number of excitons from the active junction.

## CONCLUSIONS

To summarize, we have identified the electrical signature of excitonic presence in GaInP/AlGaInP MQW and in MBE grown InGaAs/GaAs QD laser diodes. We have used bias activated differential capacitance (*fdC/df*) measurements of light emitting junctions to probe the presence of dipolar contribution of excitons within these quantum confined structures. To explain the variation of bias activated junction capacitance with applied modulation frequency, we conceived a phenomenological rate equation considering the effects of bias on the dynamics of various rate limited processes. Inversion of slope in Arrhenious like diagrams indicates to 'negative activation energy' which occurs due to the presence of stable, steady state population of the exciton like transitional bound states. In fact, this concept of 'negative activation energy' is actually borrowed here from descriptions of physical chemistry and biophysics, where it was routinely used to explain the decrease in rates of some reactions with increasing temperatures. This quintessential interdisciplinary concept is then further applied to identify steady state populations of excitonic bound states in quantum confined laser diodes using impedance spectroscopy. In MQW samples, the presence of excitons is favored only above a certain injection level, thus creating enough number of electron-hole pairs within the active region, needed to form a steady state excitonic population. The calculated average energy of this bound transitional state matches well with the binding energy of the weakly confined excitons in both the laser diodes. We emphasize that such occurrence of ‘negative activation energy’ was not observed[37] in ordinary Silicon diodes which do not show electroluminescence and perhaps are non-excitonic in nature.

Further increase in charge injection suppresses the differential capacitive response, which shows gradual Mott transition of exciton states into electron hole plasma in GaInP/AlGaInP MQW. This all-electrical, steady state description of excitonic presence and its subsequent suppression is fully supported by standard optical spectroscopic signatures of excitonic Mott transition. It may lead to innovative applications based on efficient electrical control of excitonic devices. Work is going on to use such generic approach to fully study specific analytical details of dipolar contributions of excitons through dielectric measurements.

Moreover, one may apply similar analysis to investigate the presence of condensed excitonic phases at low temperatures and to other rate limited transitions involving any bound states/quasiparticles in various materials/device structures intended for advanced functionality including ‘excitonics’. Specifically, there is increasing consensus about the role of dark excitons[54,55] in connection with still elusive excitonic BEC. It has been repeatedly advised that standard optical tools, which indirectly[22,56,57] study such condensations by looking only at photons coming out once excitonic quasi particles decay, might actually overlook the BEC all together. At this point, we speculate that our dielectric experiments dealing with frequency dependent dynamic signatures of dipolar excitons will be able to throw more light in that direction as these tools will also be sensitive to presence of the dark excitons.

## ACKNOWLEDGEMENTS

Authors wish to thank IISER-Pune for startup funding of the laboratory infrastructure as well as the grant # SR/S2/CMP-72/2012 from Dept. of Science and Technology (DST), India. AB, KB are thankful to DST, India for Inspire Fellowship and CSIR, India for Research Fellowship respectively. MH acknowledges support from the UK Engineering and Physical Sciences Research Council.

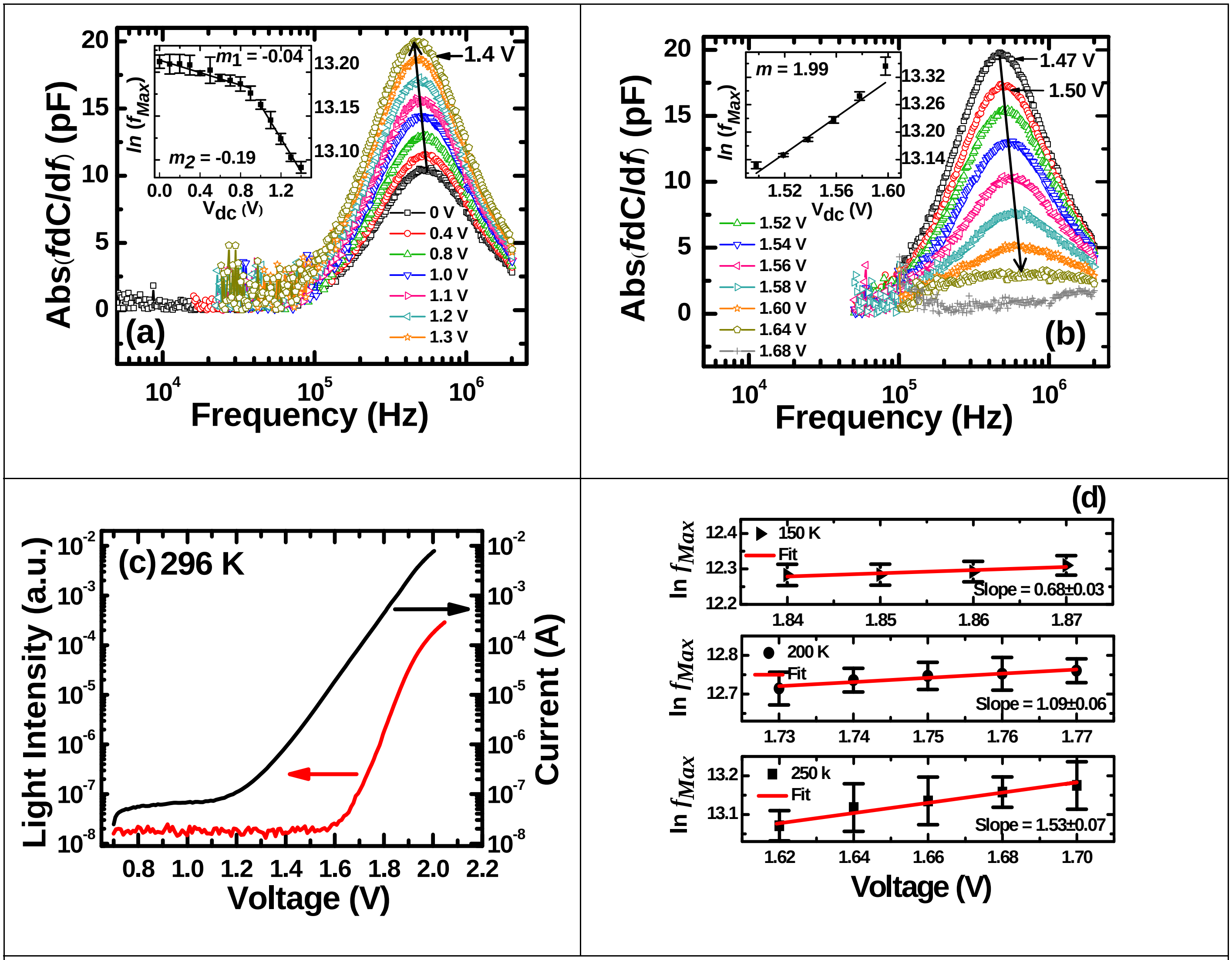

FIG. 1. Voltage activated, steady state, small signal, dielectric response of GaInP/AlGaInP MQW lasers diodes - (a) before and (b) after the onset of EL intensity around 1.5 V at 296 K. Inset in Fig. 1a shows usual linear variation of ln($f_{Max}$) with applied bias ($V_{dc}$) with slopes $m_1$ and $m_2$. This follows our proposed phenomenological rate equation (2), where $f_{Max}$ coincides with position of maximum transition rate R* in frequency domain. Inset in Fig. 1b shows large positive slopes (m) indicating ‘negative activation energy’ due to the formation of bound states which we later relate with excitons. (c) Measured current and EL intensity plotted as a function of applied bias voltage where both the values begin to increase ~1.1 V and ~1.5 V respectively. (d) Temperature dependent variation of the slopes in ln($f_{Max}$) vs voltage bias ($V_{dc}$) plots below room temperature.

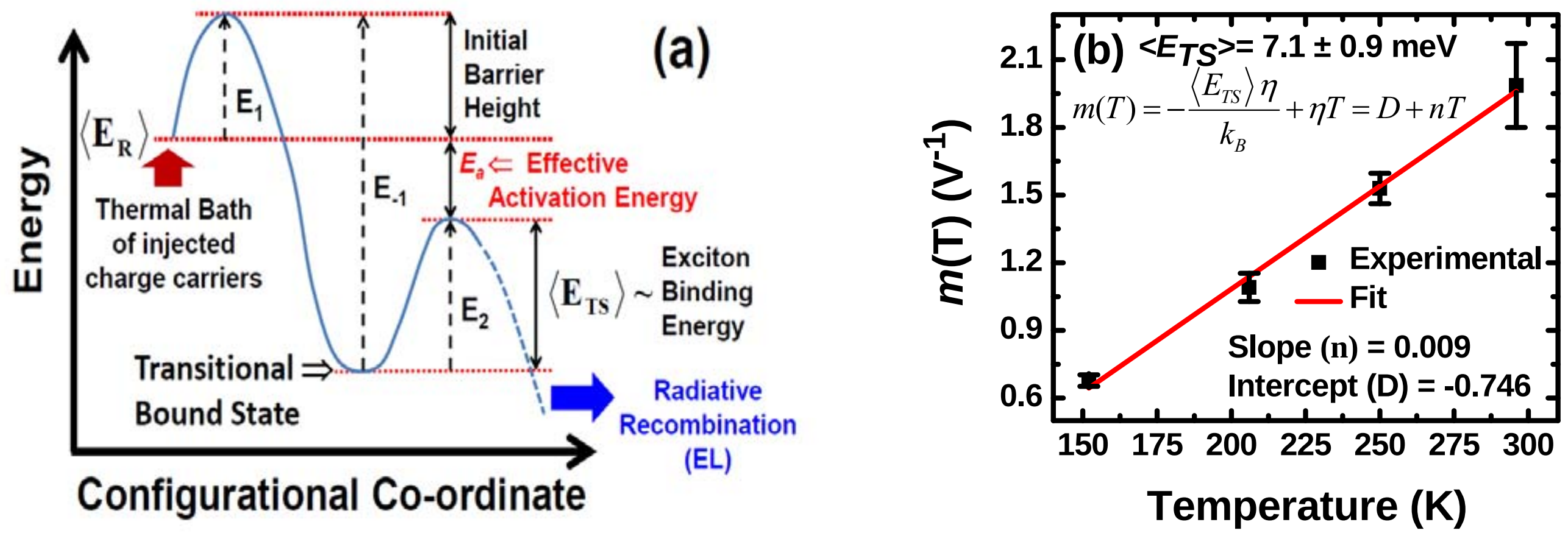

FIG. 2. | (a) Schematic of the bias activated process of EL in the configurational free energy diagram following the transition state theory of activated response. Injected charge carriers undergo a transition through an intermediate bound state and then recombine radiatively. Presence of such thermodynamical transition state (such that $E_{-1} > E_1+E_2$) leads to overall transition rate having effective 'negative activation energy' ($E_a$) as in equation (6). (b) Temperature variation of slopes *m(T)* obtained from ln($f_{Max}$) vs $V_{dc}$ plots for GaInP/AlGaInP MQW laser diodes as sown in Figs. 1b and 1d. Thermodynamical estimate of average energy of this bound transitional state (using equation (7)) is 7.1±0.9 meV. This agrees well with excitonic binding energy of weakly confined GaInP/AlGaInP MQW structures. That is one of the important reasons why we assign this stable, bound, transitional state in Fig. 2a with excitons.

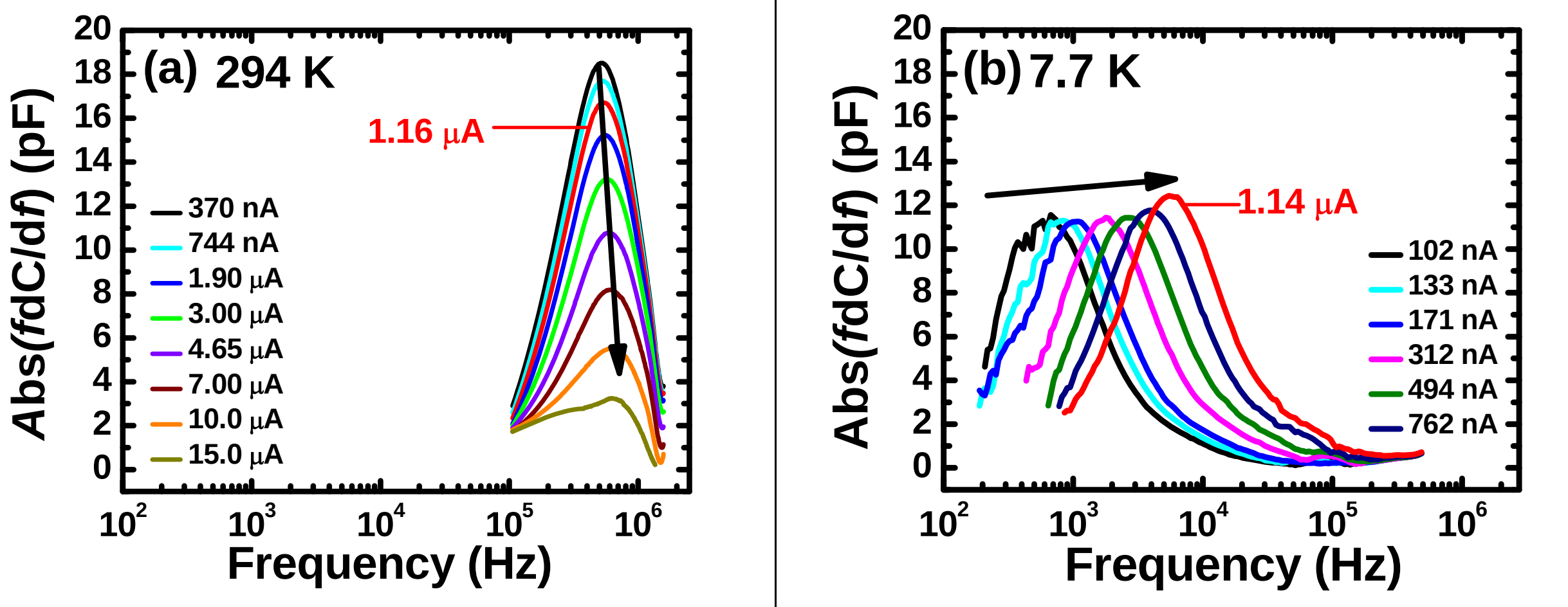

FIG. 3. | Variation of *fdC/df* with *f* for GaInP/AlGaInP MQW lasers diodes at - (a) 294 K and (b) 7.7 K. Horizontal and vertical axes of these plots are kept identical. We compare approximately similar range of applied injection currents (say ~1.15 μA, red legend) instead of bias voltages to ensure identical number of injected carriers. The $f_{Max}$ values shift toward higher frequencies with increasing injections in both the figures. This clearly indicates the presence of 'negative activation energy' regimes at both temperatures. Greater radiative recombination at 7.7 K reduces the number of available steady state excitons and thereby shrinks the *fdC/df* peak heights as compared with that of the room temperature under almost similar injections. Arrows represent frequency shift of peak position with increasing bias currents.

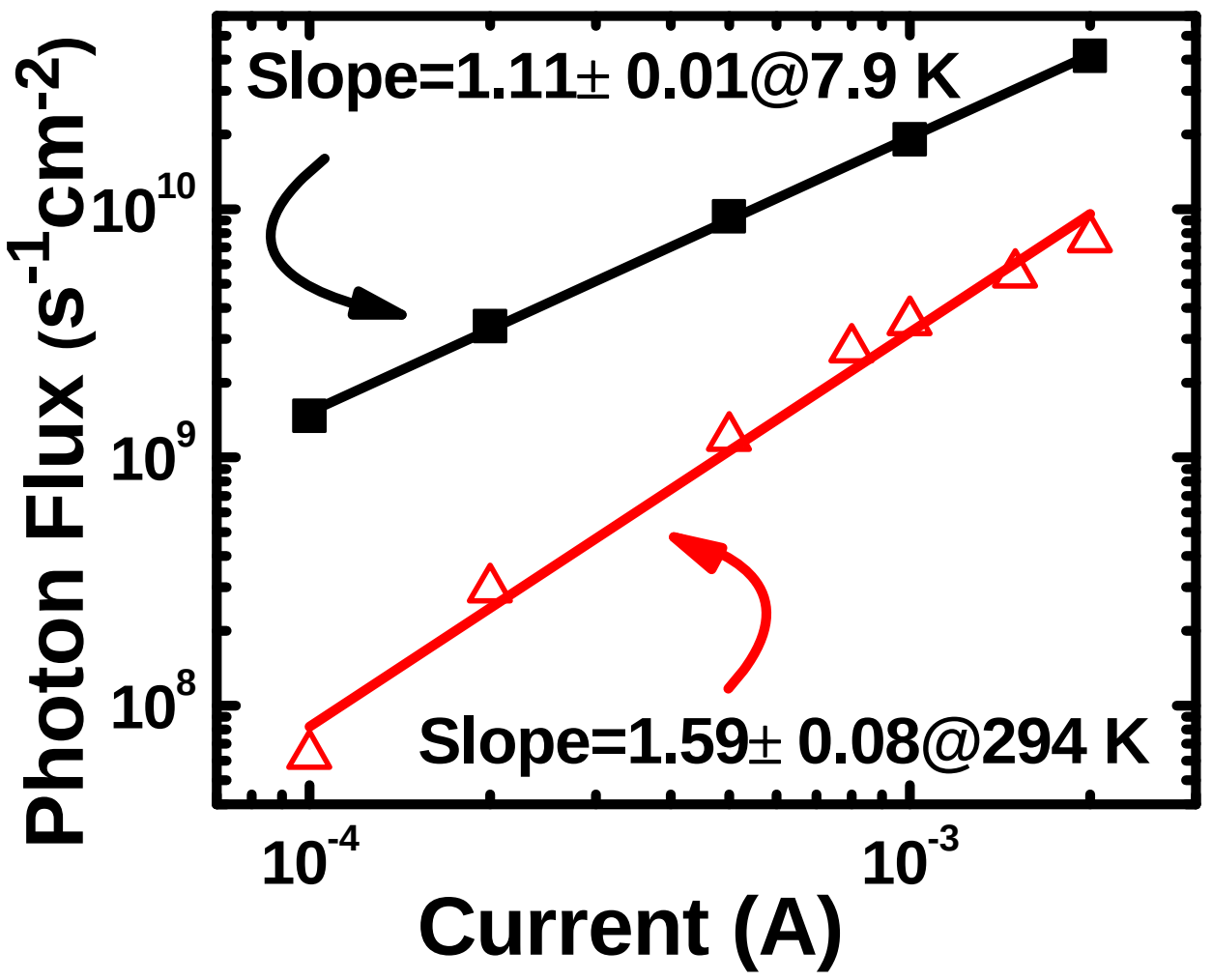

FIG. 4. | Variation of photon flux as a function of injection current at two different temperatures for GaInP/AlGaInP MQW lasers diodes. Log-log scale plots clearly imply power law behavior of photon flux at similar ranges of injection currents. Estimated power law exponent is ~1.11 at 7.9 K, which signifies the predominance of excitonic recombination. It also explains why *fdC/df* signature of excitonic Mott transition is absent around 7.7 K due to the stronger presence of excitons (Fig. 3b). This exponent further changes to 1.59 at 294 K, which indicates the higher probability for free carrier recombinations.

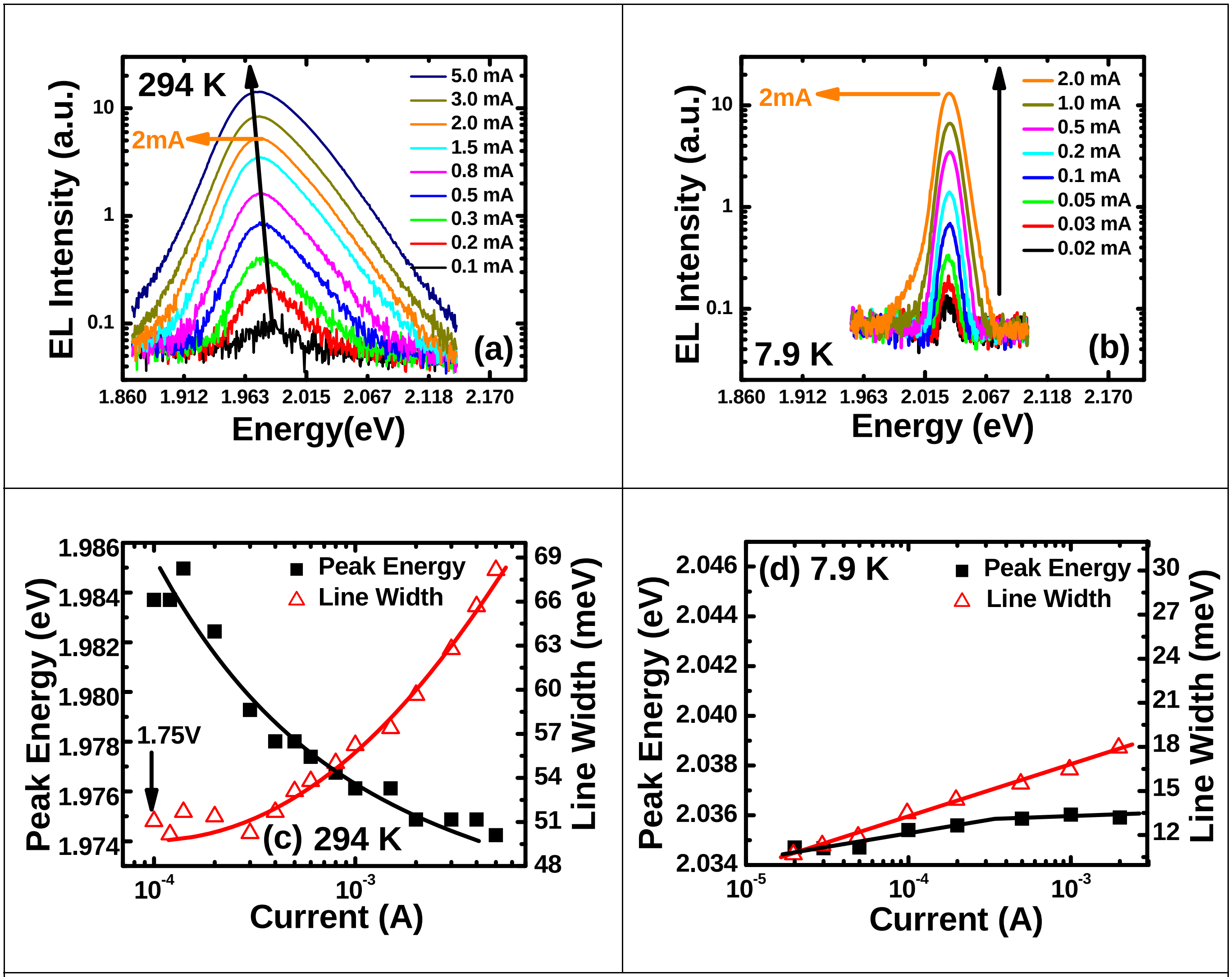

FIG. 5. | EL spectra at 294 K and 7.9 K for GaInP/AlGaInP MQW lasers diode. These are compared within partly similar ranges of applied bias currents (say 2 mA) instead of bias voltages to ensure identical number of injected carriers. Vertical arrows indicate increasing bias currents and relative shift in peak positions. (a) EL peaks slowly red shift with increasing injection at 294 K. More and more exponentially shaped high energy tails of EL spectra point to the presence of significant non-degenerate population of electron and holes for increasing injection currents as expected from Fig. 4. (b) At a much lower temperature (7.9 K), EL spectra show well formed resonantly sharp and symmetric peaks. This is unlike flatter, asymmetric and exponentially shaped high energy spectral tails observed at 294 K. Systematic variations of estimated line widths and peak energies are shown in Figs. (c) and (d) for

two temperatures. Relatively well shaped spectra measured at injection levels ≥ 1.75 V are considered for such estimations. Lines are drawn for easy visualizations only. At these high levels of injections, presence of excitonic Mott transition is apparent at 294 K. The peak energy red shifts significantly and line width broadens, which indicate the dominance of free carrier recombinations. In contrast, comparatively smaller variation of peak energies and line width are observed at 7.9 K. This is due to the stronger presence of excitonic recombination around 7.9 K in accordance with Figs. 3b and 4.

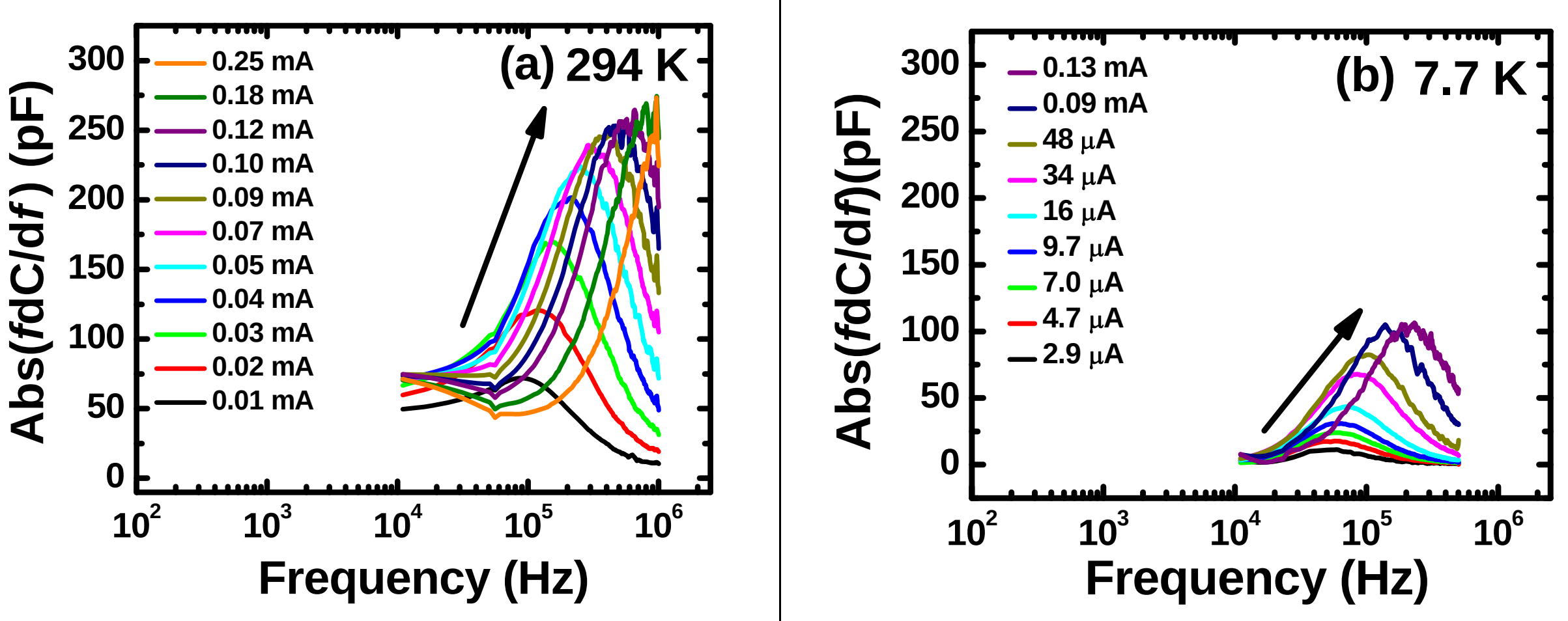

FIG. 6. | Differential capacitance response at 294 K and 7.7 K for InGaAs/GaAs QD lasers diodes. These are compared within partly similar ranges of applied bias currents instead of bias voltage to ensure identical number of injected carriers. (a) Pronounced *fdC/df* peaks for InGaAs/GaAs quantum dot laser sample are seen around room temperature. (b) Due to the higher radiative recombinations, these *fdC/df* peaks also reduces substantially at 7.7 K as compared to the plots shown in Fig. 3 for GaInP/AlGaInP MQW lasers in nearly similar range of injection currents. Frequency at the maximum activation ($f_{Max}$) shifts to higher side indicating characteristic 'negative activation energy' for excitonic presence. Tilted arrows represent frequency shift of peak position with increasing bias currents which is contrary to usual defect related changes seen by impedance spectroscopy. Frequency ranges are kept similar to that of Fig. 3 for comparison. No electrical signature of Mott transition is observed in these QD devices.

## References:

[1]I. Pelant and J.Valenta, *Luminescence Spectroscopy of Semiconductors*, Ch.7 (Oxford University Press, New York, 2012).

[2]C. F. Klingshirn, *Semiconductor Optics,* (Springer-Verlag, Berlin Heidelberg, 4th ed. 2012).

[3] R. J. Elliott, Phys. Rev. **108** 1384 (1957).

[4]R. S. Knox, and N. Inchauspe, Phys. Rev. **116** 1093 (1959).

[5]L. V. Butov, C. W. Lai, A. L.Ivanov, A. Gossard, and D.S. Chemla, Nature **417** 47 (2002).

[6]J. P. Eisenstein, and A. H. MacDonald, Nature **432** 691 (2004).

[7]A. A. High et al. Nature **483** 584 (2012).

[8]S. A.Moskalenko, and D. W. Snoke, *Bose-Einstein Condensation of Excitons and Biexcitons: and Coherent Nonlinear Optics with Excitons*, (Cambridge University Press, Cambridge 2005).

[9]E. Hanamura, and H. Haug, Phys. Rep. **33** 209 (1977).

[10]H. Deng, H. Haug, and Y. Yamamoto, Rev. Mod. Phys. **82** 1489 (2010).

[11]J. Kasprzak *et al.* Nature **443** 409 (2006).

[12]A. Kavokin, Phys. Status Solidi B **247** 1898 (2010).

[13]B. Deveaud, Annu. Rev. Condens. Matter Phys. **6** 155 (2015).

[14]V. Timofeev, and D. Sanvitto, *Exciton Polaritons in Microcavities.* (Springer-Verlag, Berlin Heidelberg, 2012).

[15]F. P.Laussy, A.V. Kavokin, and I. A. Shelykh, Phys. Rev. Lett. **104** 106402 (2010).

[16]F. P. Laussy, T.Taylor, I. A. Shelykh, and A. V. Kavokin, J. of Nanophotonics **6** 064502 (2012).

[17]L. V. Keldysh, Contemp. Phys. **27** 395 (1986).

[18]A. E. Almand-Hunter et al. Nature **506** 471 (2014).

[19]A. Imamoglu, R. J. Ram, S. Pau, and Y. Yamamoto, Phys. Rev. A **53** 4250 (1996).

[20]H.Deng, G. Weihs, D.Snoke, J. Bloch, and Y. Yamamoto, Proc. Natl. Acad. Sci. U.S.A. **100** 15318 (2003).

[21]P. Bhattacharya et al. Phys. Rev. Lett. **112** 236802 (2014).

[22]T.Byrnes, N. Y. Kim, and Y. Yamamoto, Nat. Phys. **10** 803 (2014).

[23]N. F.Mott, Rev. Mod. Phys. **40** 677 (1968).

[24]D. S. Chemla, and D. A. B. Miller, J. Opt. Soc. Am. B **2**, 1155 (1985). See Fig. 6.

[25] P.Dawson, G. Duggan, H. I.Ralph, and K. Woodbridge, Phys. Rev. B **28** 7381 (1983).

[26]M. Noriyasu, and K. Fujiwara, Appl. Phys. Lett. **97** 031103 (2010).

[27]D. A. B.Miller, *Optical Physics of Quantum Wells - Quantum Dynamics of Simple Systems*, (ed Oppo, G. L. et al.) 239-266 (Institute of Physics, London, 1996);

Available at: http://www-ee.stanford.edu/~dabm/181.pdf (Date of access: 01/10/2015)

[28]C. Klingshirn, R.Hauschild, J. Fallert, and H. Kalt, Phys. Rev. B **75** 115203 (2007).

[29]M. Z. Baten et al. Sci. Rep. **5** 11915 (2015).

[30]K. Bansal, and S. Datta, J. Appl. Phys. **110** 114509 (2011).

[31]K. Bansal, Phys Status Solidi C. **10** 593 (2013).

[32]K. Bansal, and S. Datta, Appl. Phys. Lett. **102** 053508 (2013).

[33]G. L. Miller, D. V. Lang, and L. C. Kimerling, Ann. Rev. Mater. Sci. **7** 377 (1977).

[34]T. Walter, R. Herberholz, C. Mülle,r and H. W. Schock, J. Appl. Phy*s*. **80** 4411 (1996).

[35]P. P. Boix et al. *Appl. Phys. Lett.* **95** 233302-3 (2009).

[36]A. Polimeni et al. Appl. Phys. Lett. 73 1415 (1998).

[37]K. Bansal, M. Henini, M. S. Alshammari, and S. Datta, Appl. Phys. Lett. **105** 123503 (2014).

[38]M. Mozurkewich, and S. W. Benson, J. Phys. Chem. **88** 6429 (1984).

[39]J. L. Muench, J. Kruuv, and J. R. Lepock, Cryobiology **33** 253 (1996).

[40]J. R. Alvarez-Idaboy, N. Mora-Diez, and A. Vivier-Bunge, J. Am. Chem. Soc. **122** 3715 (2000).

[41]A. Menon, and N. Sathyamurthy, J. Phys. Chem. **85** 1021 (1981).

[42]M. D. Sturge, Phys. Rev. **127** 768 (1962).

[43]C. T. H. F. Liedenbaum, A. Valster, A. L. G. J. Severens, and G. W't. Hooft, Appl. Phys. Lett. **57** 2698 (1990).

[44]J. Shao, A. Dörnen, R. Winterhoff, and F. Scholz, J. Appl. Phys. **91** 2553 (2002)

[45] J. E. Fouquet, and A. E. Siegman, Appl. Phys. Lett. **46** 280 (1985).

[46]X. Zhongying et al. Solid State Commun. **61** 707 (1987).

[47]G. T. Dang, H. Kanbe, and M. Taniwaki, J. Appl. Phys. **106** 093523 (2009).

[48]L. Kappei, J. Szczytko, F. Morier-Genoud, and B. Deveaud, Phys. Rev. Lett. **94** 147403 (2005).

[49]G.Rossbach et al. Phys. Rev. *B* **90** 201308R (2014).

[50]M. Stern, V. Garmider, V. Umansky, and I. Bar-Joseph, Phys. Rev. Lett. **100** 256402 (2008).

[51]Ioffe Institute's Online Electronic Archive, *New Semiconductor Materials (NSM), Characteristics and Properties, $Ga_xIn_{1-x}P$ Basic Parameters*. (1998-2001)

Available at: http://www.ioffe.rssi.ru/SVA/NSM/Semicond/GaInP/basic.html (Date of access: 01/10/2015).

[52]G. Wang et al. Appl. Phys. Lett. **64** 2815 (1994).

[53]S. L. Tyana et al. Solid State Comm. **117** 649 (2001).

[54]M. Combescot, O. Betbeder-Matibet, and R. Combescot, Phys. Rev. Lett. **99** 176403 (2007).

[55]E. R.Schmidgall et al. Appl. Phys. Lett. **106** 193101 (2015).

[56]D. Snoke, *Phys. Stat. Sol.(b)* **238**, 389 (2003).

[57]D. Snoke, *Indirect Excitons in Coupled Quantum Well*, Final Report. (2014)

Available at: http://www.osti.gov/scitech/servlets/purl/1141286 (Date of access: 10/02/2016).

## ADDITIONAL INFORMATION:

### Competing financial interests

The author(s) declare no competing financial interests.

### Author contributions statements

A.B., K.B. and S.D. designed the experiments; A.B. and K.B. performed these experiments and collected the data; quantum dot samples were fabricated and supplied by M.H and M.A.; A.B and K.B. wrote the main manuscript text and prepared figures; S.D. and M.H have also contributed to the interpretation and analysis of the data. All authors reviewed the manuscript.